\documentclass[twocolumn,secnumarabic,amssymb, nobibnotes, aps, prd]{revtex4-1}
\usepackage{graphicx}

\begin{abstract}
\begin{description}
\item[Abstract]The $Q$-value of the double-electron capture in $^{108}$Cd has been determined to be (272.04 $\pm$ 0.55) keV in a direct measurement with the double-Penning trap mass spectrometer TRIGA-TRAP. Based on this result a resonant enhancement of the decay rate of $^{108}$Cd is excluded. We have confirmed the double-beta transition $Q$-values of $^{106}$Cd and $^{110}$Pd recently measured with the Penning-trap mass spectrometers SHIPTRAP and ISOLTRAP, respectively. Furthermore, the atomic masses of the involved nuclides ($^{106, 108, 110}$Cd, $^{106, 108, 110}$Pd) have been directly linked to the atomic mass standard.
\end{description}
\end{abstract}

\begin{document}

\title{Direct mass measurements of cadmium and palladium isotopes and their double-beta transition $Q$-values}%

\author{C.~Smorra$^{1,2,3}$}
\email[Present authors email adress: ]{smorrac@uni-mainz.de}%
\author{T.~Beyer$^{1,3}$}
\author{K.~Blaum$^{1}$}
\author{M.~Block$^{4}$}
\author{Ch.E.~D\"ullmann$^{2,4,5}$}
\author{K.~Eberhardt$^{2}$}
\author{M.~Eibach$^{2,3}$}
\author{S.~Eliseev$^{1}$}
\author{Sz.~Nagy$^{1,4}$}
\author{W.~N\"ortersh\"auser$^{2,4}$}
\author{D.~Renisch$^{1,2}$}

\affiliation{$^1$ Max-Planck-Institut f\"ur Kernphysik, Saupfercheckweg 1, D-69117 Heidelberg}
\affiliation{$^2$ Institut f\"ur Kernchemie, Johannes Gutenberg-Universit\"at, Fritz-Strassmann-Weg 2, D-55128 Mainz}
\affiliation{$^3$ Fakult\"at f\"ur Physik und Astronomie, Ruprecht-Karls-Universit\"at, Philosophenweg 12, D-69120 Heidelberg}
\affiliation{$^4$ GSI Helmholtzzentrum f\"ur Schwerionenforschung, Planckstra\ss e 1, D-64291 Darmstadt}
\affiliation{$^5$ Helmholtz-Institut Mainz, Johannes Gutenberg-Universit\"at, D-55099 Mainz}
\maketitle

An open question in neutrino physics is whether neutrinos are their own antiparticles, i.e.~of Majorana type. This is adressed by experiments which aim to detect neutrinoless double-beta decay (0$\nu\beta\beta$) or neutrinoless double-electron capture (0$\nu\epsilon\epsilon$), where both emitted neutrinos would annihilate, thus proving the Majorana character of neutrinos \cite{RevModPhys.80.481}. If existing at all, these processes are very difficult to observe due to their long half-lives. In the case of double-beta decay experiments, nuclides with a large $Q$-value are favored, where $Q$ is the mass difference of the mother and daughter nuclide. This is due to the fact that the decay rates of the two-neutrino double-beta decay (2$\nu\beta\beta$) and the neutrinoless process (0$\nu\beta\beta$) scale with $Q^{11}$ and $Q^{5}$, respectively \cite{PTP.66.1739}. In general, the half-life of the double-electron capture is expected to be longer than that of the double-beta decay. However, if the two states in the mother and daughter nuclides undergoing a double-electron capture are degenerate in energy, the decay rate $\lambda_{\epsilon\epsilon}$ of its neutrinoless mode could be enhanced by several orders of magnitude \cite{Voloshin82,Bernabeu198315}, so that the search for the 0$\nu\epsilon\epsilon$-process is also feasible, as for example in the case of $^{152}$Gd \cite{PhysRevLett.106.052504}. The 0$\nu\beta\beta$-decay rate $\lambda_{\epsilon\epsilon}$ is proportional to
\begin{equation}
\lambda_{\epsilon\epsilon} \propto \frac{\Gamma}{(Q - B_{2h} - E_\gamma)^2 + \Gamma^2 / 4},
\label{EQ_1}
\end{equation}
where $\Gamma$ denotes the sum of the widths of the two electron-hole state and the excited nuclear state, $B_{2h}$ is the energy of the two-electron-hole state, and $E_\gamma$ is the nuclear excitation energy of the daughter nuclide. A resonant enhancement of the double-electron capture occurs if $(Q - B_{2h} - E_\gamma)$ is in the order of $\Gamma$ or smaller \cite{Voloshin82,Bernabeu198315}. The identification of resonant transitions is often limited by the uncertainty of the $Q$-value. Therefore, precise measurements of the $Q$-values with an uncertainty of a few hundered eV are necessary to identify nuclides with resonantly enhanced double-electron capture transitions.

Penning traps are nowadays the best tool for this purpose \cite{Klaus20061}, as recently shown e.g.~for the nuclides $^{76}$Ge and $^{74}$Se \cite{PhysRevC.81.032501}, $^{152}$Gd~\cite{PhysRevLett.106.052504}, $^{156}$Dy \cite{PhysRevC.84.012501}, $^{116}$Cd and $^{130}$Te \cite{Rahaman2011412}, and $^{136}$Ce~\cite{Kolhinen2011116}. In this work we have determined the atomic masses of six nuclides, $^{106,108,110}$Cd and $^{106,108,110}$Pd, and determined the $Q$-value of one double-beta decay and two double-electron capture processes.

The measurements were performed with TRIGA-TRAP \cite{Ketelaer2008162} using a laser ablation ion source \cite{0953-4075-42-15-154028}, and two Penning traps in a 7-T superconducting magnet to perform the mass measurements via the time-of-flight ion-cyclotron-resonance (TOF-ICR) method \cite{springerlink:10.1007/BF01414243}.

The laser ion source uses a rotatable target holder, in order to be able to switch between the cadmium and palladium samples for the $Q$-value measurements, and glassy carbon for carbon-cluster reference ions, which are required for the direct mass measurements. The target selection ensured that there was no isobaric contamination present during the measurements. Non-isobaric contaminations were removed in the first cylindric Penning trap by the mass-selective buffer-gas cleaning technique \cite{Savard1991247}. Subsequently, the cyclotron frequency measurement is performed in the second hyperbolic trap.

The $Q$-value was determined by measuring the cyclotron-frequency ratio $r = \nu_m / \nu_d$ in the Penning trap, where $\nu_m$
and $\nu_d$ are the cyclotron frequencies of the mother and daughter nuclides, respectively. The $Q$-value is given by:
\begin{equation}
Q = M_m - M_d = (M_m - m_e) ( 1 - r ),
\label{EQ_2}
\end{equation}
where $M_m$ and $M_d$ are the atomic masses of the mother and the daughter nuclides, respectively, and $m_e$ the electron mass. The cyclotron-frequency ratios $r$ for the $Q$-value determination were obtained by recording alternately the cyclotron frequencies of cadmium and palladium ions of the same atomic mass number $A$ under identical measurement conditions. The cyclotron frequency of the ions was measured using the TOF-ICR method using a Ramsey-excitation scheme \cite{Martin2007122, George2007110} with two excitation pulses of 100 ms and a waiting time of 800 ms in between. 

\begin{table*}[!ht]
\begin{ruledtabular}
\caption{Results of the $Q$-value measurements. The average cyclotron-frequency ratios $r$ of the mother nuclides to the daughter nuclides are given, as well as the $Q$-value derived from this value using eq.~(\ref{EQ_2}). The mass measurement based $Q$-value is obtained from the mass difference of the measurements in Table \ref{TABLE_2}. The $Q$-values from the latest Penning-trap measurements and from the atomic-mass evaluation (AME) 2003 \cite{Audi2003337} are given for comparison. }
\begin{tabular}{cccccc} 
Transition              & Cyclotron         & Direct measurement & Mass measurement   & Latest Penning trap                    & AME 2003 \cite{Audi2003337}     \\ 
                        & frequency ratio   & $Q$-value / keV  & $Q$-value / keV    & $Q$-value / keV                        & $Q$-value / keV \\ \hline
$^{106}$Cd-$^{106}$Pd   &  0.9999718705(57) & 2775.01(0.56)    & 2775.8(2.5)        & 2775.39(0.10)\cite{PhysRevC.84.028501} & 2770(7) \\
$^{108}$Cd-$^{108}$Pd   &  0.9999972934(55) &  272.04(0.55)    &  271.4(1.9)        & --                                     & 272(6) \\
$^{110}$Pd-$^{110}$Cd   & 0.9999802903(114) & 2017.8(1.2)    & 2017.3(2.5)        & 2017.85(0.64)\cite{FINK_TBP}           & 2004(11) \\ 
\end{tabular}
\label{TABLE_1}
\end{ruledtabular}
\end{table*}

\begin{figure*}[!ht]
\centering
\includegraphics[scale=0.915]{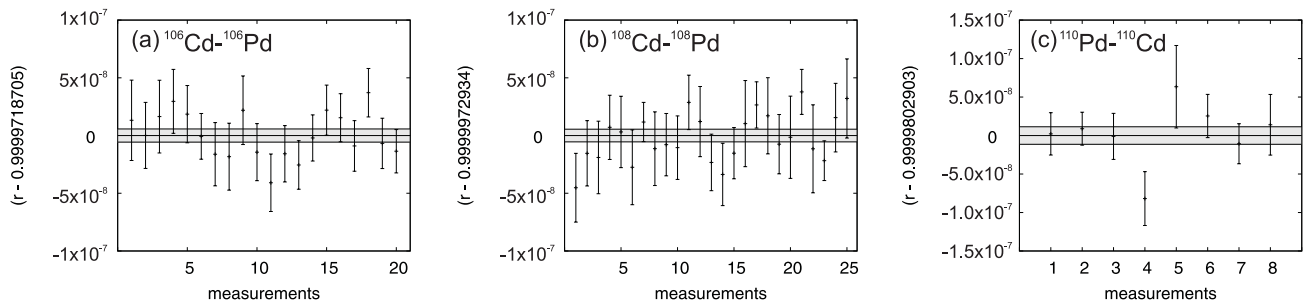}
\caption[cyclotron-frequency ratios $r$ of the $Q$-value measurements]{Individual cyclotron-frequency ratios $r$ measured for the $Q$-value determination for $^{106}$Cd-$^{106}$Pd (a), $^{108}$Cd-$^{108}$Pd (b), and $^{110}$Pd-$^{110}$Cd (c). The central lines show the average cyclotron-frequency ratios, the grey bands are their 1-$\sigma$ uncertainties. }
\label{FIG_1}
\end{figure*}

Three individual cyclotron-frequency measurements are combined for one cyclotron-frequency-ratio determination. The first and the last measurements are performed with one ion species and the cyclotron frequency is interpolated to the time of the cyclotron-frequency measurement of the other ion species \cite{springerlink:10.1140/epjd/e2002-00222-0}. The data evaluation is described in detail in \cite{springerlink:10.1140/epjd/e2010-00092-9}. In these measurements the recording time of one ion cyclotron resonance was kept below one hour in order to limit the systematic uncertainty due to non-linear magnetic-field drifts. For the evaluation only events with less than six detected ions were considered and a count-rate class analysis has been performed as described in \cite{springerlink:10.1140/epjd/e2002-00222-0} in order to account for frequency shifts due to multiple trapped ions. 

The $Q$-values can also be calculated by taking the mass differences $M_m$-$M_d$ from the mass measurements. It is advantageous to measure directly the cyclotron-frequency ratio of the mother and daughter nuclide, since in measurements of mass doublets mass-dependent uncertainties are generally negligible ($\delta r / r < 10^{-11}$) and only one cyclotron-frequency ratio has to be determined for the $Q$-value measurement. However, the direct cyclotron-frequency ratio measurements do not reveal the absolute atomic masses $M$ of the nuclides. They are obtained by determining the cyclotron-frequency ratio $r$ of the ion of interest to a carbon cluster ion and by using:
\begin{equation}
M = r (M_{\mathrm{ref}} - m_e) + m_e,
\label{EQ_3}
\end{equation}
where $M_{\mathrm{ref}}$ is the mass of the reference ion. In this case, the cyclotron-frequency ratios were recorded in alternate measurements between carbon-cluster reference ions and the cadmium or the palladium isotope. The cyclotron-frequency ratios were obtained analogous to the $Q$-value measurements. A systematic shift depending on the mass difference between the ion of interest and the reference mass needs to be considered for the cyclotron-frequency ratios of ions with different mass number $A$, which is caused by a slight misalignment between the electric and magnetic field of the second Penning trap \cite{springerlink:10.1140/epjd/e2010-00092-9}. This shift was determined to be (4.2 $\pm$ 6.6) 10$^{-10} \times \Delta m / \mathrm{u} $ by analysing the cyclotron-frequency ratios of C$_9^+$ to carbon-cluster ions from C$_{7}^+$ to C$_{11}^+$.  The cyclotron-frequency ratio was corrected by the shift, and the uncertainty of the shift was quadratically added to the statistical uncertainty of the cyclotron-frequency ratios. 

\begin{table*}[!ht]
\begin{ruledtabular}
\caption{Results of the direct mass measurements. The corrected cyclotron-frequency ratios $r$ of the ions of interest to the carbon-cluster reference ions are given, as well as the relative uncertainties of the measurement $\delta r / r$, the mass excesses $\Delta$ calculated from each cyclotron-frequency ratio and the weighted-mean values of the mass excesses $\Delta_{\mathrm{avg}}$, and the mass excesses $\Delta_{\mathrm{lit}}$ listed in the AME 2003 \cite{Audi2003337}.}
\begin{tabular}{cccccccc} 
Nuclide     & Reference ion &  $r$            &  $\delta r / r$        & $\Delta$ / keV & $\Delta_{\mathrm{avg}}$ / keV &$\Delta_{\mathrm{lit}}$ / keV \\ \hline
$^{106}$Cd	& C$^+_9$    & 0.980615266(16) & 1.6 $\times$ 10$^{-8}$ & -87132.6(1.6)  & -87133.5 (1.9) & -87132 (6) \\
           	& C$^+_8$    & 1.103192822(35) & 3.2 $\times$ 10$^{-8}$ & -87137.2(3.1)  &                &            \\ \hline

$^{106}$Pd	& C$^+_9$    & 0.980587670(18) & 1.9 $\times$ 10$^{-8}$ & -89908.8(1.8)  & -89909.3 (1.7) & -89902 (4) \\
           	& C$^+_8$    & 1.103161788(49) & 4.4 $\times$ 10$^{-8}$ & -89912.5(4.4)  &                &            \\ \hline

$^{108}$Cd	& C$^+_9$    & 0.999112789(16) & 1.6 $\times$ 10$^{-8}$ & -89254.2(1.6)  & -89254.0 (1.5) & -89252 (6) \\ 
          	& C$^+_{10}$ & 0.899201063(40) & 4.4 $\times$ 10$^{-8}$ & -89253.1(4.5)  &                &            \\ \hline

$^{108}$Pd	& C$^+_9$    & 0.999110092(12) & 1.2 $\times$ 10$^{-8}$ & -89525.5(1.2)  & -89525.4 (1.2) & -89524 (3) \\ 
          	& C$^+_{10}$ & 0.899198638(39) & 4.3 $\times$ 10$^{-8}$ & -89524.2(4.3)  &                &            \\ \hline

$^{110}$Cd	& C$^+_9$    & 1.017620488(14) & 1.4 $\times$ 10$^{-8}$ & -90352.1(1.4)  & -90350.8 (2.1) & -90353.0 (2.7) \\
          	& C$^+_{10}$ & 0.915858014(20) & 2.1 $\times$ 10$^{-8}$ & -90347.6(2.2)  &                &                \\ \hline

$^{110}$Pd	& C$^+_9$    & 1.017640557(16) & 1.6 $\times$ 10$^{-8}$ & -88333.2(1.7)  & -88333.5 (1.4) & -88349 (11) \\ 
          	& C$^+_{10}$ & 0.915876025(25) & 2.7 $\times$ 10$^{-8}$ & -88334.4(2.8)  &                &             \\ 
\end{tabular}
\label{TABLE_2}
\end{ruledtabular}
\end{table*}

Table \ref{TABLE_1} shows the results of the direct $Q$-value measurements. The individual cyclotron-frequency ratios of these measurements are shown in Fig.~\ref{FIG_1}. The mass measurement $Q$-values listed in Table \ref{TABLE_1} are calculated by using the differences from the mass measurements listed in Table \ref{TABLE_2}. Both measurement methods are in agreement within their uncertainty. The larger uncertainties of the mass measurement $Q$-values are due to less statistics compared to the direct $Q$-value measurements.

The isotope $^{106}$Cd is an interesting candidate for double-$\beta^+$ decay and double-electron capture experiments due to the large $Q$-value and its comparably high natural abundance \cite{berglund2011isotopic}. Recently, the upper limits on the half-lives of the double-$\beta^+$ and the electron-capture processes have been investigated \cite{1748-0221-6-08-P08011, Belli.arXiv.2011}. We measured the $Q$-value of this transition (see Table \ref{TABLE_1}). The result is in agreement with a recent measurement by SHIPTRAP \cite{PhysRevC.84.028501}. A degeneracy in energy was reported \cite{PhysRevC.84.028501}, but due to the different parity of the ground state of the mother and the excited nuclear state of the daughter nuclide and the low probability of electron capture from the L$_3$-atomic orbital, the isotope $^{106}$Cd is not suited for the search of the $0\nu\epsilon\epsilon$-process \cite{Krivoruchenko2011140}. The mass value of $^{106}$Pd differs by 7 keV from the AME 2003 value \cite{Audi2003337}, whereas the mass value of $^{106}$Cd agrees with the AME 2003 value.

The $Q$-value of double-electron capture in $^{108}$Cd was determined to be $272.04(0.55)$ keV in the first direct measurement of this transition. Thereby, the literature value of the AME 2003 was confirmed and the precision improved by more than an order of magnitude. The measured mass values of $^{108}$Cd and $^{108}$Pd as listed in Table \ref{TABLE_2} agree within one standard deviation with previous measurements \cite{PhysRev.132.1673, Firestone.IAEA.2007, PhysRevC.21.65, PhysRevC.21.1667}. In addition, the uncertainties of the mass values were improved by a factor of 4 and 2.5, respectively. The resonance condition of the double-electron capture is not fulfilled for this nuclide, since the lowest excited state in $^{108}$Pd has an excitation energy of 433.938(5) keV \cite{Jean2000135}. 

We have investigated the double-beta decay $Q$-value of $^{110}$Pd, which showed a discrepancy of 16 keV from the AME 2003 value. The discrepancy is due to the mass value of $^{110}$Pd, which is off by 16 keV. The mass of $^{110}$Cd is in agreement with the AME 2003 value. Our results are in agreement with the $Q$-value and mass measurements reported from ISOLTRAP \cite{FINK_TBP}.  

In conclusion, three $Q$-values of double-beta transitions were measured. In case of $^{106}$Cd and $^{110}$Pd results from previous experiments \cite{PhysRevC.84.028501, FINK_TBP} were confirmed. The $Q$-value of the double-electron capture in $^{108}$Cd was measured with a factor of ten reduced uncertainty. The resonance condition for the double-electron capture in $^{108}$Cd is not fulfilled in this transition. Furthermore, the mass values of the six investigated nuclides were directly linked to the atomic mass standard.

The financial support by the Max-Planck Society as well as technical support of the Nuclear Chemistry Department at the University of Mainz is acknowledged. Sz. Nagy acknowledges financial support from the Alliance Program of the Helmholtz Association (HA216/EMMI).

\end{document}